\documentclass[11pt]{article}
\usepackage{amsmath, amssymb} 
\usepackage{graphicx}
\usepackage[colorlinks=true,citecolor=blue,linkcolor=black]{hyperref}

\def\x{{\bf x}}

\def\v{{\bf v}}
\def\L{{\scriptscriptstyle L}}
\def\E{{\scriptscriptstyle E}}

\def\kB{k_{\rm\scriptscriptstyle B}}
\def\coeff#1#2{{\textstyle {\frac {#1}{#2}}}}

\begin{document}

\title{\Large Temperature in relativistic fluids}
\author{\normalsize Pavel Kovtun\\
{\it\normalsize Department of Physics \& Astronomy,  University of Victoria,}\\
{\it\normalsize PO Box 1700 STN CSC, Victoria, BC,  V8W 2Y2, Canada}
}

\date{\normalsize March, 2023}

\maketitle

\begin{abstract}
\noindent
For static matter in a gravitational field, different conventions for equilibrium local temperature exist in the classic physics literature. We illustrate the difference between two popular conventions using black-body radiation in a spherically symmetric gravitational potential. Equilibrium temperatures defined by the ``Landau frame'' or ``Eckart frame'' prescriptions most commonly used in relativistic fluid dynamics do not satisfy the statistical-mechanical relation $1/T=dS/dE$.
\end{abstract}

\section{Introduction}
\begin{figure}
\centering
\includegraphics[width=0.4\textwidth]{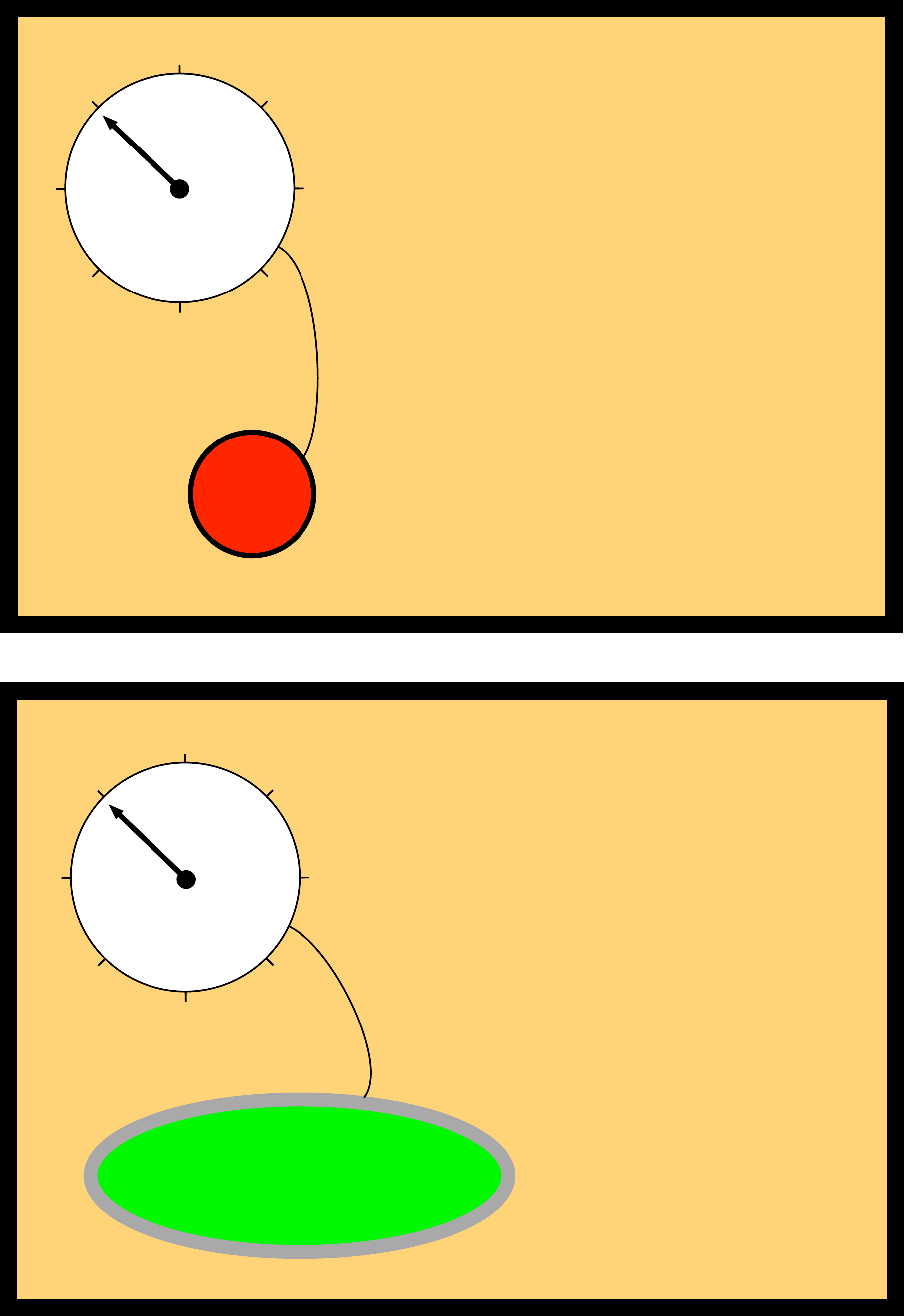}
\hspace{0.16\textwidth}
\includegraphics[width=0.4\textwidth]{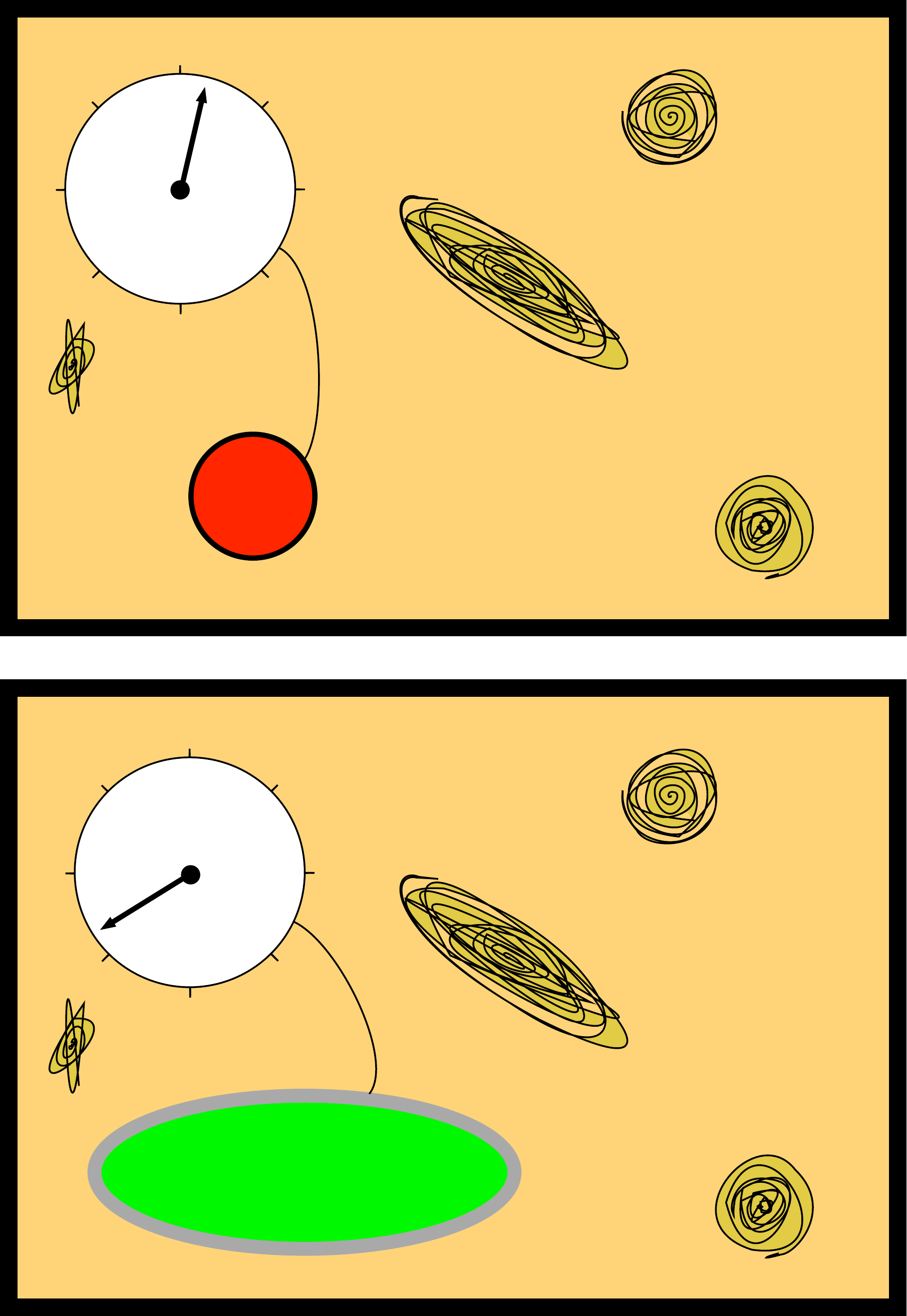}
\caption{
\label{fig:thermometers}
Left: two differently constructed but properly calibrated thermometers will show the same temperature when placed in identical equilibrium states. A thermometer is visualized as a small reservoir (depicted as a colored blob) in thermal contact with the system. The thermometer's reservoir is coupled to a gauge (the white dial) that is used to read off the temperature. Right: the thermometers that show the same temperature in identical equilibrium states will in general show different temperatures in identical non-equilibrium states.
}
\end{figure}

\noindent
Temperature is a fundamental physical quantity that characterizes matter in equilibrium~\cite{Reif, LL5}. For equilibrium matter which is spatially uniform, temperature is constant everywhere. The absolute temperature, when measured locally with a properly calibrated thermometer, is well-defined and universal: it does not depend on the type of thermometer used. In other words, two different thermometers, when used to measure temperatures of two identical systems in equilibrium, will show the same result. This is illustrated in Fig.~\ref{fig:thermometers}, left. The same universality does not hold out of equilibrium of course: two different properly calibrated thermometers, when placed in identical non-equilibrium states, will in general show different temperatures, as illustrated in Fig.~\ref{fig:thermometers}, right. This is because a thermometer is a macroscopic object with a finite size and a finite response time to the disturbances of the environment. Thus the notion of non-equilibrium, space- and time-dependent temperature $T(t,\x)$ is not universal, but rather is thermometer-dependent.%
\footnote{
In fluid dynamics, similar ambiguities also exist for the non-equilibrium velocity field $\v(t,\x)$. For example, one is free to define different non-equilibrium ``fluid velocities'' corresponding to the flow of energy, the flow of particles, the flow of entropy etc, and then formulate the fluid-dynamical equations using one's preferred choice. We will only consider temperature ambiguities here. In the example we study later in the paper (blackbody radiation in static equilibrium), different definitions of the fluid velocity agree, but different definitions of temperature do not.
}
This degree of non-universality will be small for non-equilibrium states whose macroscopic parameters change slowly in space and time.

Thus it appears that the locally measured out-of-equilibrium temperature is not universal, but the equilibrium temperature is. However, a moment's thought reveals that this statement may not necessarily be true in the presence of gravity. For a system subject to an external gravitational field, the static equilibrium state does not have to be spatially uniform, as for example happens for a gas that forms the atmosphere of a massive planet. In a non-uniform equilibrium state, it is possible that even the equilibrium temperature is not universal, but rather depends on the size and shape of the thermometer. The difference in the temperature readings between two properly calibrated thermometers will be of order of at least one spatial derivative of the gravitational field.

For definiteness, let us take our matter to be a (relativistic) fluid subject to a static external gravitational field described by a time-independent metric $g_{\mu\nu}(\x)$. As argued by Tolman a long time ago~\cite{Tolman:1930zza, Tolman}, the local temperature $T(\x)$ of the fluid in equilibrium is
\begin{equation}
\label{eq:Tolman-T}
  T(\x) = \frac{T_0}{\sqrt{-g_{00}(\x)}}\,,
\end{equation}
with some constant $T_0$. [Conventions: the metric is mostly plus.] The Tolman temperature (\ref{eq:Tolman-T}) is widely viewed as ``the'' temperature of the fluid in equilibrium in a gravitational field, and is usually discussed in textbooks that talk about both thermal physics and gravitation, see e.g.~\cite{LL5}, \S 27, or \cite{Frolov-Zelnikov}, sec.~9.6. Given the above discussion, one can ask: is the locally measured equilibrium temperature in the presence of gravity universal, or is it dependent on one's choice of convention? 

In fact, different inequivalent definitions of equilibrium temperature for matter in a gravitational field have  existed in the classic physics literature for quite some time. A case in point is the {\em Course of Theoretical Physics} by L.~D.~Landau and E.~M.~Lifshitz. In volume~V, {\em Statistical Physics}, the authors argue that the equilibrium temperature is indeed given by the Tolman's expression~(\ref{eq:Tolman-T}), see Ref.~\cite{LL5}, \S~27. Less explicit is the argument in volume~VI, {\em Fluid Mechanics}, see~\cite{LL6}, \S~136, where a general procedure for writing down the equations of relativistic fluid dynamics is proposed. One might think that the equilibrium (hydrostatic) solutions of the fluid-dynamical equations obtained by using the Landau-Lifshitz procedure of Ref.~\cite{LL6} will give rise to the equilibrium temperature~(\ref{eq:Tolman-T}). As we will see shortly, this is {\em not} the case. In other words, the hydrostatic $T(\x)$ determined by the procedure of volume~VI does not agree with the thermodynamic $T(\x)$ advocated in volume~V. The fact that there is a difference between these two temperatures is well known to the handful of people to whom it is well known. However, for a student of statistical physics and relativity, the discrepancy may cause confusion. Given the popularity of the Landau-Lifshitz prescription \cite{LL6} in modern relativistic fluid dynamics~\cite{Romatschke:2017ejr}, it is worthwhile to shed more light on the details, and to illustrate this difference with a simple pedagogical example. This is the goal of the present article.%
\footnote{
  The temperature differences we discuss here are not related to the Hawking radiation, vacuum selection, or the presence of horizons. Neither are they related to one's preferred choice of the equilibrium Killing vector in stationary non-static states. Throughout the following discussion, the metric is fixed and non-dynamical, and Einstein's equations are never used. 
}
We will compare the two popular temperature definitions, and will argue that the definition of temperature which maintains Eq.~(\ref{eq:Tolman-T}) is preferred, while the definition of temperature advocated in Ref.~\cite{LL6} is best avoided.

Before we explore this difference quantitatively, let us first recall how the equilibrium temperature (\ref{eq:Tolman-T}) is derived. An argument familiar from basic physics comes from Maxwell's velocity distribution in equilibrium~\cite{Reif}. For a non-relativistic particle of mass $m$ with gravitational potential energy $mgz$, the phase-space probability distribution is proportional to $\exp (-[\frac12mv^2 + mgz]/\kB T_0)$, with constant $T_0$. Now, the gravitational potential energy in the field of a spherical body of mass $M$ and radius $R$ is really $-\frac{GMm}{R} + mgz + \dots$, where $G$ is Newton's constant. The relativistic kinetic energy is $E_{\rm kin} = mc^2 + \frac12 mv^2 + \dots$, hence the probability is proportional to 
\begin{align}
  e^{-\frac{E}{\kB T_0}} = e^{-\left[ mc^2 + \frac12 mv^2 - \frac{GMm}{R} + mgz + O(1/c^2) \right]/\kB T_0} = e^{-\frac{E_{\rm kin}}{\kB T_0}\sqrt{-g_{00}}} = e^{-\frac{E_{\rm kin}}{\kB T}}\,,
\end{align}
where $T=T_0/\sqrt{-g_{00}}$ is the height-dependent temperature, $g_{00} = -1+\frac{2GM}{c^2r}$ is the metric component encoding the gravitational potential, and $r=R+z$, with $z\ll R$. From now on, we will set the Boltzmann constant $\kB$ and the speed of light $c$ to one. 

Tolman's derivation~\cite{Tolman} assumes that the equilibrium matter such as the black-body radiation can be modeled by the ``perfect fluid'' energy-momentum tensor,
\begin{equation}
\label{eq:Tmn0}
  T^{\mu\nu}_{(0)} = \epsilon u^\mu u^\nu + p (g^{\mu\nu} {+} u^\mu u^\nu)\,,
\end{equation}
where $\epsilon(T)$ is the energy density, $p(T)$ is the pressure, and $u^\mu$ is the fluid velocity, $u^2=-1$. The conservation law $\nabla_{\!\mu} T^{\mu\nu}=0$ then implies the equilibrium profile (\ref{eq:Tolman-T}). The argument in Ref.~\cite{LL5} is different: the temperature is defined by $1/T = \partial S/\partial E$, where $S$ is the entropy, and $E$ is the energy of the equilibrium matter. Given that the energy of a small volume of matter as measured by a local observer is redshifted by a factor of $\sqrt{-g_{00}}$, and assuming that the entropy of that volume of matter is unchanged by the gravitational field, one arrives at Eq.~(\ref{eq:Tolman-T}). There are similar derivations such as~\cite{Balazs:1963uu}, see e.g.~Ref.~\cite{Santiago:2018kds} for a recent pedagogical discussion. 
It is then clear that the assumptions leading to Eq.~(\ref{eq:Tolman-T}) can only hold if the gravitational field varies arbitrarily slowly in space. It is conceivable that the statistical-mechanical temperature $(\partial S/\partial E)^{-1}$ will be modified in curved space as the small volume of matter that is used to define $T$ senses curvature. Similarly, the perfect fluid approximation~(\ref{eq:Tmn0}) is only true to leading order in the derivatives of the metric: the general expression for the energy-momentum tensor of equilibrium matter will have extra terms determined by the Riemann curvature, such as $R^{\mu\nu}$, $u_\alpha R^{\alpha(\mu\nu)\beta}u_\beta$ etc, in addition to those in Eq.~(\ref{eq:Tmn0}). 
This is the gravitational analogue of polarization, when the density of the electric charge in the presence of a non-uniform electromagnetic potential $(\phi, {\bf A})$ receives contributions proportional to the derivatives of the potential  (electric and magnetic fields). Similarly, the energy-momentum tensor of matter in the presence of a non-uniform metric receives contributions proportional to the derivatives of the metric (Riemann curvature tensor). Analogous to the electric susceptibility which measures charge density fluctuations in thermal equilibrium, the gravitational susceptibility measures stress fluctuations in thermal equilibrium. Such equilibrium stress fluctuations are non-zero even in flat space.
Going beyond the ``perfect-fluid'' approximation of (\ref{eq:Tmn0}), one indeed finds ambiguities in what the equilibrium temperature might be. Let us now discuss this in detail.

\section{Prescriptions}
\noindent
Let the static external metric vary on length scales much longer than the microscopic length scale of the fluid (such as the thermal de Broglie wavelength $\lambda_{\rm th}$). The energy-momentum tensor then has the form of a gradient expansion,
\begin{equation}
\label{eq:Tsum}
  T^{\mu\nu} = T^{\mu\nu}_{(0)} + \sum_{n\geqslant1} T^{\mu\nu}_{(n)}\,,
\end{equation}
where $T^{\mu\nu}_{(n)}$ schematically denotes symmetric tensors made out of $T,u^\alpha, g_{\alpha\beta}$ that have $n$ derivatives.%
\footnote{
  We are assuming a finite correlation length. For superfluid phases where long-range correlation are mediated by gapless ``Goldstone bosons" $\varphi$, the expansion will also include derivatives of $\varphi$. We focus on normal fluids for simplicity.
}
Among the terms in this expansion, some terms will vanish in equilibrium (such as the $n=1$ terms corresponding to the shear and bulk viscosities), and some terms will be non-zero in equilibrium (such as the ``perfect fluid'' $n=0$ terms, and certain $n=2$ terms).

The ambiguities in how exactly one chooses to define $T$ and $u^\alpha$ for a given $T^{\mu\nu}$ may be fixed by various prescriptions. For example, the Landau-Lifshitz prescription~\cite{LL6} at each order in the derivative expansion demands  $T^{\mu\nu}_{(n)}u_\nu = 0$ for $n\geqslant 1$. Let us denote the temperature in the Landau-Lifshitz prescription by $T_\L$ and the fluid velocity in the Landau-Lifshitz prescription by $U^\mu$. The Landau-Lifshitz prescription amounts to defining the fluid velocity as the normalized timelike eigenvector of the energy-momentum tensor, i.e.\ $T^{\mu\nu} U_\nu  = - \epsilon\, U^\mu$, and defining the temperature by the eigenvalue $- \epsilon(T_\L)$, where the function $\epsilon(T_\L)$ is given by the equilibrium flat-space equation of state. The prescription applies to both equilibrium and non-equilibrium terms in Eq.~(\ref{eq:Tsum}) equally. When restricted to hydrostatic equilibrium, the prescription fixes a definition for the equilibrium temperature. This Landau-Lifshitz prescription is often called ``the Landau frame''.

The Landau-frame prescription for temperature is just a prescription (out of an infinite multitude of possible prescriptions) for how to extract a scalar quantity we call ``$T_\L$'' from the energy-momentum tensor $T^{\mu\nu}$. Importantly, the Landau-frame prescription for extracting ``$T_L$'' from $T^{\mu\nu}$ is formulated without trying to ensure that the quantity ``$T_L$'' is in fact consistent with the statistical-mechanical temperature in equilibrium, when going beyond the perfect-fluid approximation. As we will see later, ignoring the connection with statistical physics is dangerous: simply postulating a random prescription for how to extract from $T^{\mu\nu}$ a scalar function one chooses to label as ``temperature'' may lead to unphysical answers.

A different prescription for equilibrium quantities was advocated in Refs.~\cite{Banerjee:2012iz, Jensen:2012jh}. In order for the fluid subject to the external metric to remain in equilibrium, it is necessary that the metric has a time-like Killing vector (corresponding to the time-translation invariance), call it $V$. The equilibrium temperature is then defined by $T=T_0/\sqrt{-V^2}$, and the equilibrium fluid velocity by $u^\mu = V^\mu/\sqrt{-V^2}$ to all orders in the derivative expansion.% 
\footnote{
  For conformal matter (such as the blackbody radiation), one can relax the requirement of having a timelike Killing vector, and only require that $V$ is a timelike conformal Killing vector. As an example, in the Friedmann-Robertson-Walker metric this definition of temperature gives $T\propto 1/a(t)$, where $a(t)$ is the scale factor. This gives the temperature in a ``radiation-dominated'' universe, familiar from elementary cosmology. 
}
In other words, the prescription of \cite{Banerjee:2012iz, Jensen:2012jh} fixes the equilibrium temperature by demanding that Tolman's law~(\ref{eq:Tolman-T}) continues to hold to all orders in the derivative expansion (in equilibrium), consistent with the argument in Ref.~\cite{LL5}.  Following~\cite{Jensen:2012jh}, we shall call this equilibrium prescription ``the thermodynamic frame''. This latter prescription does not fix $T$ and $u^\alpha$ out of equilibrium. For non-equilibrium definitions, one may further choose either the Landau-Lifshitz-like prescription $T^{\mu\nu}_{(n),\textrm{non-equil.}}u_\nu = 0$, or some other prescription.

The actual energy-momentum tensor must of course remain the same, regardless of which prescription for $T$ and $u^\mu$ one chooses to adopt.

\section{Spherical hydrostatic equilibrium}
\label{sec:example}
\noindent
To be concrete, let us consider a definite physical system, the black-body radiation. For such a fluid, the energy-momentum tensor defined with the Landau-Lifshitz prescription was written down in Ref.~\cite{Baier:2007ix}, up to second order in derivatives, i.e.\ including terms up to $n=2$ in the expansion (\ref{eq:Tsum}). For fluid in equilibrium, the answer in 3+1 dimensions is
\begin{align}
\label{eq:TLL-2}
  T^{\mu\nu} =  p(T_\L) g^{\mu\nu} + 4p(T_\L) U^\mu U^\nu
  +\kappa(T_\L) \left[ R^{\langle\mu\nu\rangle} -2 U_\alpha R^{\alpha \langle\mu\nu\rangle \beta}U_\beta \right] + \lambda(T_\L) \omega^{\langle \mu}_{\ \ \alpha} \omega^{\nu\rangle \alpha} \,,
\end{align}
where $R^{\mu\nu}$ is the Ricci tensor, $R^{\alpha\mu\nu\beta}$ is the Riemann tensor, and the angular brackets denote the transverse (to $U_\mu$) symmetric traceless part of a tensor. 
The vorticity tensor is defined as $\omega^{\mu\nu} = \frac12 \Delta^{\mu\alpha} \Delta^{\nu\beta}(\nabla_\alpha U_\beta - \nabla_\beta U_\alpha)$, where $\Delta^{\mu\nu}\equiv g^{\mu\nu} + U^\mu U^\nu$ projects onto the space orthogonal to $U^\mu$, and $\nabla_{\!\alpha}$ is the covariant derivative. 
The coefficients $\kappa$ and $\lambda$ are thermodynamic susceptibilities whose values have to be found from the microscopic theory.
For  black-body radiation (ideal gas of photons in equilibrium), $p(T)=\pi^2 T^4/(45\hbar^3)$~\cite{LL5}, while for $\kappa$ one finds $\kappa(T) = T^2/(18\hbar)$ \cite{Romatschke:2009ng}. More generally, we will write $p(T)=p_0 T^4$, $\kappa(T) = \kappa_0 T^2$, as dictated by dimensional analysis for a generic radiation fluid. The coefficient $\lambda(T)$ has also been evaluated for black-body radiation~\cite{Moore:2012tc}, however it will not be needed for the discussion of non-rotating fluids below.%
\footnote{
  We emphasize that $\kappa$ and $\lambda$ describe real physical effects (gravitational and vortical susceptibilities), and as such are independent on one's prescription for $T$ and $u^\mu$. A prescription (Landau-Lifshitz or otherwise) determines where $\kappa$ and $\lambda$ appear in $T^{\mu\nu}$, but does not change the  value of $\kappa$ and $\lambda$. See also sec.~\ref{sec:translation}.
}

Expression \eqref{eq:TLL-2} does not assume any particular model of matter, nor any particular calculational technique, such as kinetic theory or holography. Rather, Eq.~\eqref{eq:TLL-2} is written on symmetry grounds, assuming that: {\it a)} the energy-momentum tensor in equilibrium is a local function of the metric and its derivatives,  {\it b)} when the curvature is small on the length scale set by the state of the fluid ($\lambda_{\rm th} \partial g \ll 1$, where $\lambda_{\rm th}$ is the thermal de Broglie wavelength), its contribution to $T^{\mu\nu}$ can be treated perturbatively as a small correction to the perfect-fluid energy-momentum tensor, {\it c)} matter has conformal symmetry (which is a symmetry of black-body radiation and other radiation fluids), {\it d)} fluid's temperature and velocity are defined by the Landau-Lifshitz prescription. For more details, see Ref.~\cite{Baier:2007ix}.

Let us now explore the consequences of Eq.~(\ref{eq:TLL-2}) for spherical non-rotating equilibrium. The metric in the static coordinates is
\begin{align}
\label{eq:s-metric}
  ds^2 = - e^{2\Phi(r)} dt^2 + \frac{dr^2}{1-2G M(r)/r} + r^2 \left(d\theta^2 + \sin^2\!\theta\, d\varphi^2 \right)\,.
\end{align}
The Schwarzschild solution corresponds to $M(r)={\rm const}$, and $e^{2\Phi(r)} = 1-\frac{2GM}{r}$, where $G$ is Newton's constant. For the radiation fluid at rest, with vanishing $U^r, U^\theta, U^\varphi$,%
\footnote{
  The fluid velocity for the fluid at rest is thus $U^\mu = V^\mu/\sqrt{-V^2}$, where $V^\mu =(1,0,0,0)$ is the timelike Killing vector, in static coordinates, reflecting the existence of equilibrium. Thus in this example the fluid velocity is the same in the ``Landau frame'', and in the ``thermodynamic frame''. The fluid temperatures in the two conventions are different though. 
}
let us parametrize the equilibrium temperature~as 
\begin{align}
\label{eq:TL}
T_\L(r) = T_0\, e^{-\Phi(r)} h(r) \,,
\end{align}
with $h(r)\to1$ at large $r$, so that $T_0$ is the temperature at infinity. The energy-momentum conservation $\nabla_{\!\mu} T^{\mu\nu} = 0$ together with the expression (\ref{eq:TLL-2}) with vanishing vorticity now give an ordinary differential equation for the function $h(r)$, whose coefficients depend on $\Phi(r)$ and $M(r)$. For the Schwarzschild background, the equation takes a simple form:
\begin{align}
\label{eq:h-eqn}
  \rho^5 h(\rho)^2 h'(\rho) + \varepsilon \left[ \rho(1{-}\rho)h'(\rho) +\coeff14 h(\rho) \right] = 0\,,
\end{align}
where $\rho\equiv r/l_S >1$, and $l_S = 2GM$ is the Schwarzschild radius. The dimensionless parameter $\varepsilon$ in Eq.~(\ref{eq:h-eqn}) reflects the effect of the gravitational susceptibility, $\varepsilon = \kappa_0/(p_0 T_0^2 l_S^2)$. For a gas of photons, $\varepsilon = 5\hbar^2/(2\pi^2 T_0^2 l_S^2)$. To get a sense of the numerical value, recall that the temperature $T_0$ of $1K$ corresponds to $\hbar/T_0$ of about 2 millimeters, while the Schwarzschild radius of the Earth is about 9 millimeters. Physically, one expects that the derivative expansion for the energy-momentum tensor of macroscopic matter in Eq.~(\ref{eq:TLL-2}) will be valid when $\varepsilon$ is small. The condition $\varepsilon\ll 1$ also ensures that the mean photon wavelength is much smaller than~$l_S$, and the macroscopic description in terms of a fluid remains valid.
\begin{figure}[t]
\centering
\includegraphics[width=0.6\textwidth]{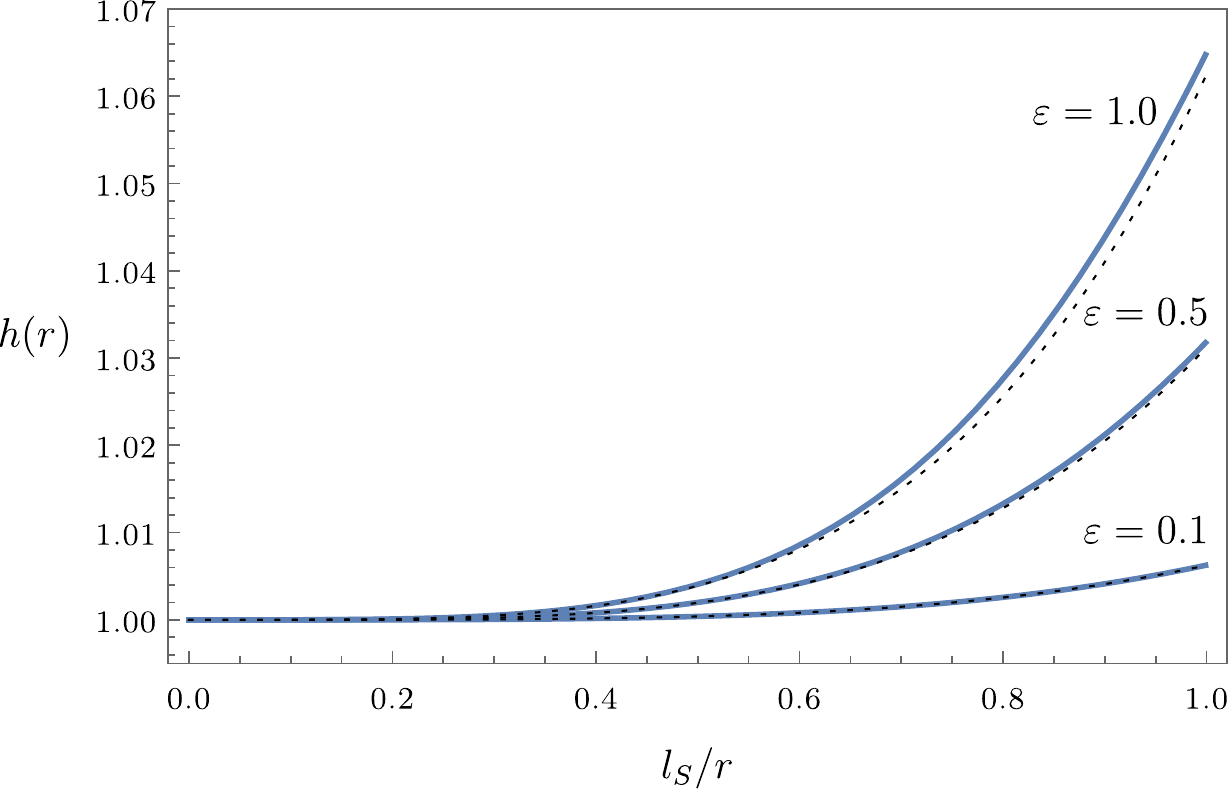}
\caption{
\label{fig:hofr}
The function $h(r)$ which determines the Landau-Lifshitz temperature through $T_\L = h(r)T_0/\sqrt{-g_{00}}$, plotted for the radiation fluid in the Schwarzschild background. The value of the parameter $\varepsilon$ is determined by the gravitational susceptibility $\kappa$ which parametrizes the leading-order response of equilibrium matter to curvature, see Eq.~(\ref{eq:TLL-2}). The solid curves are obtained by solving Eq.~(\ref{eq:h-eqn}). The leading-order approximation to the solution of Eq.~(\ref{eq:h-eqn}), $h(r) \approx 1 + \frac{\varepsilon}{16} (l_S/r)^4$ is shown by the dotted lines. The deviation of $h(r)$ from 1 shows that the hydrostatic temperature in the Landau-Lifshitz prescription does not follow Eq.~(\ref{eq:Tolman-T}).
}
\end{figure}
When $\varepsilon=0$, the solution to Eq.~(\ref{eq:h-eqn}) is $h(\rho)=1$, and Tolman's temperature (\ref{eq:Tolman-T}) is reproduced. When $\varepsilon$ is non-zero, the solutions to Eq.~(\ref{eq:h-eqn}) are plotted in Fig.~\ref{fig:hofr}. The solutions asymptote to~1 as $r\to\infty$, and stay regular as $r\to l_S$. The deviation of $h(r)$ from 1 shows that the equilibrium hydrostatic temperature in the Landau-Lifshitz prescription does not agree with Eq.~(\ref{eq:Tolman-T}). 

Also shown in Fig.~\ref{fig:hofr} is the large-$r$ solution, $h(r) = 1 + \frac{\varepsilon}{16} (l_S/r)^4 + O(1/r^7)$. It can be understood as arising from two-derivative corrections to the perfect-fluid constitutive relations. Indeed, for the Tolman temperature in the Schwarzschild background $(\partial_r T)^2/T^4 \sim (l_S/r)^4 (1/l_S^2 T_0^2)$ is the dimensionless parameter which determines the importance of two-derivative corrections. The terms beyond $O(1/r^4)$ in the solution for $h(r)$ should not be trusted, as their effect is the same as that of the higher-derivative terms not taken into account in Eq.~(\ref{eq:TLL-2}).

Near the surface of a spherical body of radius $R$, we write $r=R+z$, and find
\begin{align}
\label{eq:Tz}
  T_\L (z) = T_\L(z{=}0) \left[ 1 - \left( \frac12 \frac{l_S}{R-l_S} + \frac{\varepsilon_s}{4} \frac{R}{R-l_S} \left( \frac{l_S}{R} \right)^4 \right) \frac{z}{R} + O\left( \frac{z^2}{R^2}, \frac{z}{R}\varepsilon_s^2 \right)\right]\,,
\end{align}
where $\varepsilon_s \equiv \kappa_0/(p_0 T_L(z{=}0)^2 l_S^2)$.
The $\varepsilon_s$-independent term comes from the universal Tolman's temperature, while the $\varepsilon_s$-dependent contribution is the correction to Tolman's law due to the gravitational susceptibility $\kappa$ in the Landau-frame hydrostatics.

\section{Translating between prescriptions}
\label{sec:translation}
\noindent
To get a better handle on the discrepancy between the hydrostatic temperature (\ref{eq:TL}) in the Landau-Lifshitz prescription and Eq.~(\ref{eq:Tolman-T}), let us consider the same equilibrium configuration in the ``thermodynamic frame''~\cite{Banerjee:2012iz, Jensen:2012jh}. The latter is the prescription in which the equilibrium temperature $T\equiv T_0/\sqrt{-V^2}$ and the fluid velocity $u^\mu \equiv V^\mu/\sqrt{-V^2}$ are written in terms of the timelike Killing vector $V$, which is needed to define the equilibrium in the first place. In other words, the convention is such that Tolman's law (\ref{eq:Tolman-T}) is satisfied identically in equilibrium without any corrections, even once the gravitational susceptibilities are taken into account. The energy-momentum tensor in the ``thermodynamic frame'' is obtained by varying the effective action (equilibrium free energy for a fluid subject to external gravitational field) with respect to the metric.  The effective action does not assume any particular model of matter, nor any particular calculational technique, such as kinetic theory or holography. Rather, the effective action is written on symmetry grounds, assuming diffeomorphism invariance, locality, and derivative expansion (valid when $\lambda_{\rm th} \partial g \ll 1$). For conformal matter (such as the black-body radiation) the effective action must also be invariant under the Weyl rescaling of the metric. For more details, see Refs.~\cite{Banerjee:2012iz, Jensen:2012jh}.

Let us write the energy-momentum tensor in the thermodynamic-frame convention as
\begin{align}
\label{eq:TTF}
  T^{\mu\nu} = {\cal E} u^\mu u^\nu + {\cal P}\Delta^{\mu\nu} + {\cal Q}^\mu u^\nu + {\cal Q}^\nu u^\mu + {\cal T}^{\mu\nu} \,,
\end{align}
where ${\cal Q}^\mu$ is transverse to $u_\mu$, and ${\cal T}^{\mu\nu}$ is transverse to $u_\mu$, symmetric, and traceless. The spatial projector is again $\Delta^{\mu\nu} \equiv g^{\mu\nu} + u^\mu u^\nu$. This defines ${\cal E}$ (energy density), ${\cal P}$ (pressure), ${\cal Q}^\mu$ (energy flux) and ${\cal T}^{\mu\nu}$ (stress). For a fluid in equilibrium in a static external gravitational field, we have ${\cal E} = \epsilon(T) + f_{\cal E}$, ${\cal P} = p(T)+ f_{\cal P}$, where $f_{\cal E}, f_{\cal P}, {\cal Q}^\mu, {\cal T}^{\mu\nu}$ are $O(\partial^2)$.  Using the notation of \cite{Kovtun:2018dvd}, the energy-momentum tensor for conformal matter in external gravitational field is given by 
\begin{align}
\label{eq:E-thermo}
  & {\cal E} = \epsilon(T) + f_1 R - 6f_1 a^2 - (4f_1+f_3)\Omega^2 + 6f_1 u^\alpha R_{\alpha\beta}u^\beta + O(\partial^3)\,,\\
\label{eq:QQ}
  & {\cal Q}_\mu = (2f_1 + 4f_3) \left( \epsilon_{\mu\lambda\rho\sigma} a^\lambda u^\rho \Omega^\sigma + \Delta^\rho_\mu R_{\rho\sigma} u^\sigma \right) + O(\partial^3)\,,\\
\label{eq:TTmn}
  & {\cal T}_{\mu\nu} = (2f_3 - f_1) \Omega_{\langle\mu}\Omega_{\nu\rangle} + 4f_1 u^\alpha R_{\alpha \langle\mu\nu\rangle \beta} u^\beta - 2f_1 R_{\langle\mu\nu\rangle} + O(\partial^3)\,,
\end{align}
and ${\cal P} = \coeff13 {\cal E}$. Here $a^\mu \equiv u^\lambda \nabla_{\!\lambda} u^\mu$ is the acceleration, $\Omega^\mu \equiv \epsilon^{\mu\nu\rho\sigma}u_\nu \nabla_{\!\rho} u_\sigma$ is the vorticity vector, and $R$ is the Ricci scalar. The functions $f_1(T)$ are $f_3(T)$ are the two equilibrium susceptibilities.%
\footnote{
   For fluids whose microscopic degrees of freedom are not charged, there are in general three gravitational susceptibilities $f_1$, $f_2$, $f_3$. For radiation fluids, conformal symmetry implies $f_2 = 6f_1$. 
}
The relation to the susceptibilities which appear in \eqref{eq:TLL-2} is $\kappa(T)=-2f_1(T)$, $\lambda(T) = 2 T \partial f_1/\partial T - 8 f_3(T)$. For a gas of photons, an explicit calculation gives $f_1 = -f_3 = -T^2/(36\hbar)$. The vorticity vector is related to the vorticity tensor by $\omega^{\mu\nu} = -\frac12 \epsilon^{\mu\nu\rho\sigma}u_\rho \Omega_\sigma$. 

For equilibrium matter, its vorticity and the Ricci tensor are not independent.%
\footnote{ I thank Kristan Jensen for a discussion on this point.} 
Using the definition of the fluid velocity $u^\mu\equiv V^\mu/\sqrt{-V^2}$ and the Killing equation obeyed by $V$, it is not difficult to derive that in equilibrium
\begin{align}
  (\nabla^\mu + a^\mu) \omega_{\mu\nu} + R_{\nu\lambda} u^\lambda + (u^\alpha R_{\alpha\beta} u^\beta + \omega_{\alpha\beta} \omega^{\beta\alpha})u_\nu  = 0\,.
\end{align}
Thus for an equilibrium fluid without vorticity, the fluid velocity is an eigenvector of the Ricci tensor, and $\Delta^\rho_\mu R_{\rho\sigma} u^\sigma =0$. This is what happens in the case of a static fluid in the Schwarzschild metric (\ref{eq:s-metric}):  both the vorticity $\Omega_\mu$ and the energy flux ${\cal Q}_\mu$ vanish, the $f_3$ contributions drop out from $T^{\mu\nu}$, and the energy-momentum tensor at $O(\partial^2)$ is determined by the pressure $p(T)$ and the susceptibility $f_1(T)$.

By construction, the energy-momentum tensor \eqref{eq:TTF}--\eqref{eq:TTmn} in the thermodynamic frame is conserved in equilibrium {\em identically}. In other words, the solution to the hydrostatic conservation equations $\nabla_{\!\mu} T^{\mu\nu} = 0$ is given by $T= T_0/\sqrt{-V^2}$ and $u^\mu = V^\mu/\sqrt{-V^2}$ to all orders in the derivative expansion, which is the whole point of the ``thermodynamic frame'' construction. The function $h(r)$ of the Landau-frame hydrostatics in Eq.~(\ref{eq:TL}) thus parametrizes the deviation of the Landau-frame temperature from the thermodynamic-frame temperature.

The energy-momentum tensor (\ref{eq:TLL-2}) expressed in terms of $T_\L$ and $U^\mu$ (Landau frame), and the energy-momentum tensor (\ref{eq:TTF})--(\ref{eq:TTmn}) expressed in terms of $T$ and $u^\mu$ (thermodynamic frame) is one and the same $T^{\mu\nu}$. In an arbitrary ``primed'' frame, one can write
\begin{align}
  T^{\mu\nu} = {\cal E}' u'^\mu u'^\nu + {\cal P}' \Delta'^{\mu\nu} + {\cal Q}'^\mu u'^\nu + {\cal Q}'^\nu u'^\mu + {\cal T}'^{\mu\nu} \,,
\end{align}
with $\Delta'^{\mu\nu}=g^{\mu\nu} + u'^\mu u'^\nu$. The two equilibrium expressions of $T^{\mu\nu}$ are converted to each other by the transformation
$u'^\mu = u^\mu + \delta u^\mu$, $T' = T + \delta T$, where $\delta u^\mu$, $\delta T$ are $O(\partial^2)$, explicitly
\begin{align}
\label{eq:frame-transformation}
  \delta u^\mu = \frac{{\cal Q}^\mu - {\cal Q}'^\mu}{\epsilon+p}\,,\ \ \ \ 
  \delta T = \frac{f_{\cal E} - f'_{\cal E}}{\partial \epsilon/\partial T}\,.
\end{align}
Taking the unprimed frame to be the thermodynamic frame, and the primed frame to be the Landau frame, we have ${\cal Q}'^\mu = 0$, $f'_{\cal E} = 0$, hence $\delta u^\mu = {\cal Q^\mu}/(\epsilon+p)$, and $\delta T = f_{\cal E}/(\partial\epsilon/\partial T)$. Further, in the static background (\ref{eq:s-metric}) the energy current (\ref{eq:QQ}) vanishes, as does the vorticity~$\Omega^\mu$. Hence to $O(\partial^2)$ we have for conformal fluids $U^\mu = u^\mu$, and $T_\L = T + \delta T$, with 
\begin{align}
\label{eq:dTT}
  \frac{\delta T}{T} = \frac{f_1}{12p} \left(  R - 6  a^2 + 6  u^\alpha R_{\alpha\beta}u^\beta \right)\,.
\end{align}
For the Schwarzschild background, we have $a^2 = l_S^2/(4r^4) (1-l_S/r)^{-1}$, while both $R$ and $u^\alpha R_{\alpha\beta}u^\beta$ vanish. One thus finds from Eq.~(\ref{eq:dTT}) that the $O(\partial^2)$ transformation from the thermodynamic frame to the Landau-Lifshitz frame for conformal matter in the Schwarzschild background is given by
\begin{align}
\label{eq:deltaT}
  \frac{\delta T}{T} = \frac{\varepsilon}{16} \frac{l_S^4}{r^4}\,,
\end{align}
where $\varepsilon = \kappa_0/(p_0 T_0^2 l_S^2)$ as before. Remembering that $T_\L(r) = T(r) h(r)$, the change $\delta T$ is related to $h(r)$ by $\delta T/T = h(r) - 1$. We thus see that the leading-order transformation $\delta T$ from the thermodynamic frame to the Landau frame gives rise to the leading-order behavior of $h(r)$, shown in Fig.~\ref{fig:hofr} by dotted lines, as it should be. This shows very explicitly that the deviation of $T_\L(r)$ from Tolman's law~(\ref{eq:Tolman-T}) is a consequence of the Landau-Lifshitz hydrodynamic prescription, inherited by the Landau-Lifshitz hydrostatics. The deviation can be eliminated by converting to the thermodynamic prescription where Tolman's law is preserved.

\section{Discussion}
\noindent
Returning to the discussion in the Introduction, one may ask which temperature convention is ``better''. It seems clear that the thermodynamic convention implicitly adopted in~\cite{LL5} (in which the Tolman law stays uncorrected) is universal, while the convention implicitly adopted in~\cite{LL6} is not. In the latter case, the equilibrium temperature in the Landau-Lifshitz convention will depend on the substance whose temperature is being measured.  Indeed, Eq.~(\ref{eq:deltaT}) shows that the correction to the Tolman law in the Landau-Lifshitz convention of~\cite{LL6} depends on $\kappa_0$, $p_0$ which are different for different substances. In other words,  {\em according to the ``Landau frame'' convention of~\cite{LL6}, two different substances which have exactly the same temperature in the absence of gravity will develop different temperatures when subject to exactly the same gravitational potential}. On the other hand, the thermodynamic-convention result \eqref{eq:Tolman-T} is agnostic to the detailed microscopic nature of matter whose temperature is being measured.

\begin{figure}[t]
\centering
\includegraphics[width=0.6\textwidth]{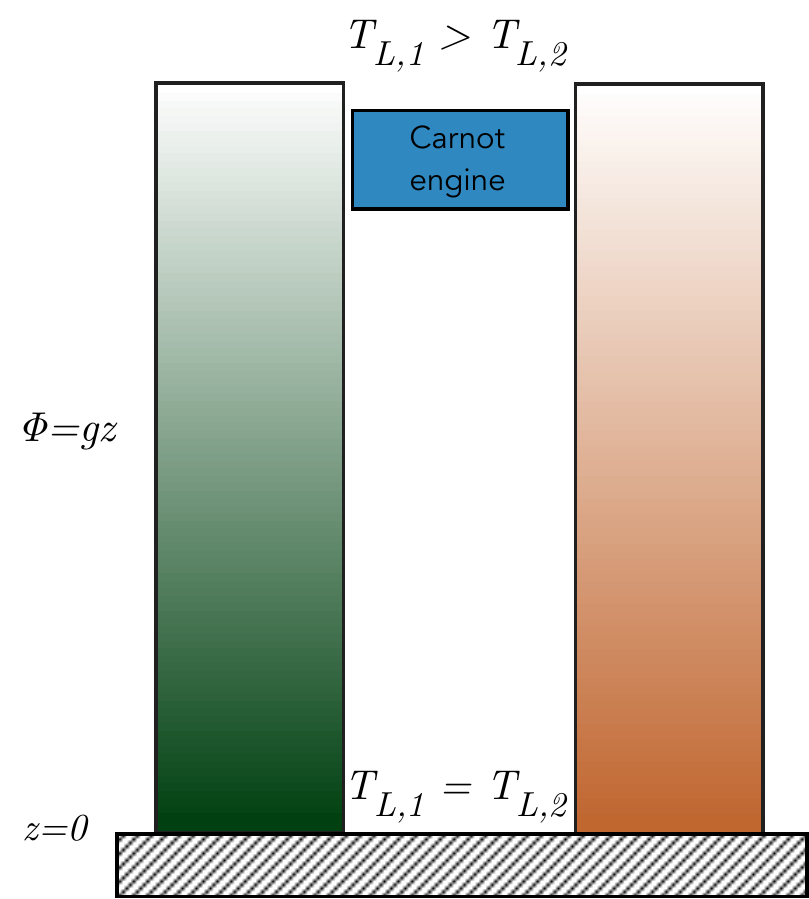}
\caption{
\label{fig:Maxwell-columns}
Maxwell's argument demonstrating why temperature gradient in a gravitational potential can not depend on the substance~\cite{Maxwell}. Two columns filled with different substances are subject to the same gravitational potential $\Phi = gz$, where $g$ is the gravitational acceleration. The columns are connected by a conducting plate at the bottom, so that at $z=0$ the temperatures are equal. If the temperatures of the two columns are not the same at high $z$, one can run a heat engine using the temperature difference between the tops of the two columns. As the engine runs, both columns will cool, while maintaining different temperatures on top. Eventually, all the energy of the two-column system will be converted into work, in violation of the second law of thermodynamics. 
}
\end{figure}

Non-universality of the Landau-frame convention is disconcerting: substance-dependence is not what we expect of a sensible physical temperature. One may recall Maxwell's argument: in {\em The Theory of Heat}~\cite{Maxwell}, Ch.~XXII, Maxwell argues that a possible temperature gradient in a gravitational potential can not be substance-dependent, otherwise one can drive a heat engine using the temperature difference between the two substances, and violate the second law of thermodynamics. This is illustrated in Fig.~\ref{fig:Maxwell-columns}. As one can see from Eq.~\eqref{eq:Tz}, the Landau-frame temperature gradient is indeed substance-dependent because $\kappa_0/p_0$ is in general different for different substances. So, does the Landau-Lifshitz temperature convention of~\cite{LL6} violate the second law of thermodynamics? 

Imagine connecting the tops of the two columns in Fig.~\ref{fig:Maxwell-columns}, so that the columns can exchange heat. The two Landau-frame temperatures on top are different, but the condition of no-heat-flow is not $T_{\L,1} = T_{\L,2}$, but rather $dS_1/dE_1 = dS_2/dE_2$, where $S_a$  and $E_a$ are the entropy and energy of the two volumes matter in heat contact, located at the same potential. In order to save the Landau-Lifshitz convention of \cite{LL6} from violating the second law, it appears that one must have $1/T_\L \neq dS/dE$ in the presence of gravity. This is consistent with the original argument in \cite{LL5}: the temperature defined by $dS/dE$ is the Tolman temperature $T$, while the Landau-frame temperature differs from $T$, as illustrated in Sec.~\ref{sec:example}.

The convention-dependence of equilibrium temperature in curved space may seem unintuitive, but ultimately it arises because the energy-momentum tensor of equilibrium matter in general depends not just on the background metric, but also on its derivatives. This dependence is parametrized through susceptibilities such as $\kappa$ in Eq.~(\ref{eq:TLL-2}), which are gravitational analogues of the electric and magnetic susceptibilities. Different conventions for $T$ and $u^\lambda$ determine where exactly the susceptibilities will appear when $T^{\mu\nu}$ is expressed in terms of the chosen $T$ and $u^\lambda$, but the $T^{\mu\nu}$ itself is of course the same, regardless of our preference for $T$ and $u^\lambda$. 

The fact that the energy-momentum tensor of an ideal quantum gas in thermal equilibrium depends on curvature has been known for many years from calculations of effective actions in thermal field theory, see e.g.~\cite{Nakazawa:1984zq, Dowker:1988jw, Gribosky:1988yk}. Gravitational susceptibilities were identified in the context of relativistic fluid dynamics in~\cite{Baier:2007ix, Bhattacharyya:2008jc, Jensen:2011xb}, and  connected to the equilibrium free energy of fluids in curved space in~\cite{Banerjee:2012iz, Jensen:2012jh}. While the direct experimental observation of gravitational susceptibilities is difficult, they can be in principle computed analytically or numerically. There are more recent results for ideal quantum gases \cite{Romatschke:2009ng, Moore:2010bu, Moore:2012tc, Megias:2014mba, Buzzegoli:2017cqy, Kovtun:2018dvd}, as well as for interacting theories including lattice QCD \cite{Philipsen:2013nea}, the O(N) model \cite{Romatschke:2019gck}, unitary Fermi gas~\cite{Lawrence:2022vwa}, and the ${\cal N}=4$ supersymmetric Yang-Mills theory~\cite{Baier:2007ix, Bhattacharyya:2008jc, Buchel:2008bz, Arnold:2011ja}. As expected, the actual values of the susceptibilities depend on the microscopic degrees of freedom and their interactions. 

We have focused on the Landau frame for definiteness when discussing temperature conventions, but it's worth mentioning another popular convention for relativistic fluid dynamics, the ``Eckart frame''~\cite{PhysRev.58.919, weinberg:1972}. In this convention, the fluid velocity is aligned with the particle number flux, while the temperature is defined in the same way as in the Landau frame, $T^{\mu\nu} U_{\mu,\E} U_{\nu,\E} = \epsilon(T_\E)$. Eckart's convention is not relevant for our example of the black-body radiation because there is no conserved particle number current. For other substances where Eckart's convention is relevant, the hydrostatic Eckart-frame temperature will suffer from the same non-universality in curved space as the hydrostatic Landau-frame temperature.

Both the Landau-Lifshitz and Eckart conventions were proposed before the gravitational susceptibilities were discovered (theoretically), and the non-universality of the corresponding temperature conventions wasn't explored at the time. In order to minimize potential confusion, it seems best to avoid using either Landau-frame or Eckart-frame temperature conventions when discussing hydrostatics beyond perfect fluids, and instead use the thermodynamic frame where Tolman's law is preserved. The Landau-frame or Eckart-frame prescriptions can then be used on top of the thermodynamic frame in order to simplify out-of-equilibrium contributions to the constitutive relations of hydrodynamics. 

Finally, we have focused on the non-universality of the Landau-frame temperature caused by the gravitational field. A similar substance-dependence for the Landau-frame temperature will also emerge in a rotating fluid without gravity, as is evident from the transformation formula \eqref{eq:frame-transformation} given the vorticity-dependent $f_{\cal E}\equiv {\cal E}-\epsilon$ in Eq.~\eqref{eq:E-thermo}. Thus, we similarly expect that the Landau-frame temperature will not be equal to $(d S/d E)^{-1}$ for rotating fluids.

\subsubsection*{Acknowledgements}
This work was supported in part by the NSERC of Canada.

\bibliographystyle{JHEP}
\bibliography{hydro-general-biblio}

\providecommand{\href}[2]{#2}\begingroup\raggedright\begin{thebibliography}{10}

\bibitem{Reif}
F.~Reif, \emph{Fundamentals of Statistical and Thermal Physics}. McGraw-Hill,
  1965.

\bibitem{LL5}
L.~D. Landau and E.~M. Lifshitz, \emph{Statistical Physics, Part I}. Pergamon,
  1980.

\bibitem{Tolman:1930zza}
R.~C. Tolman, \emph{{On the Weight of Heat and Thermal Equilibrium in General
  Relativity}}, \href{https://doi.org/10.1103/PhysRev.35.904}{\emph{Phys. Rev.}
  {\bfseries 35} (1930) 904}.

\bibitem{Tolman}
R.~C. Tolman, \emph{Relativity, Thermodynamics, and Cosmology}. Oxford, 1934.

\bibitem{Frolov-Zelnikov}
V.~Frolov and A.~Zelnikov, \emph{Introduction to Black Hole Physics}. Oxford,
  2011.

\bibitem{LL6}
L.~D. Landau and E.~M. Lifshitz, \emph{Fluid Mechanics}. Pergamon, 1987.

\bibitem{Romatschke:2017ejr}
P.~Romatschke and U.~Romatschke, \emph{{Relativistic Fluid Dynamics In and Out
  of Equilibrium}}. Cambridge University Press, 2019,
  \href{https://doi.org/10.1017/9781108651998}{10.1017/9781108651998},
  [\href{https://arxiv.org/abs/1712.05815}{{\ttfamily 1712.05815}}].

\bibitem{Balazs:1963uu}
N.~L. Balazs and J.~M. Dawson, \emph{On thermodynamic equilibrium in a
  gravitational field},
  \href{https://doi.org/https://doi.org/10.1016/0031-8914(65)90089-3}{\emph{Physica}
  {\bfseries 31} (1965) 222}.

\bibitem{Santiago:2018kds}
J.~Santiago and M.~Visser, \emph{{Tolman temperature gradients in a
  gravitational field}},
  \href{https://doi.org/10.1088/1361-6404/aaff1c}{\emph{Eur. J. Phys.}
  {\bfseries 40} (2019) 025604}
  [\href{https://arxiv.org/abs/1803.04106}{{\ttfamily 1803.04106}}].

\bibitem{Banerjee:2012iz}
N.~Banerjee, J.~Bhattacharya, S.~Bhattacharyya, S.~Jain, S.~Minwalla and
  T.~Sharma, \emph{{Constraints on Fluid Dynamics from Equilibrium Partition
  Functions}}, \href{https://doi.org/10.1007/JHEP09(2012)046}{\emph{JHEP}
  {\bfseries 09} (2012) 046} [\href{https://arxiv.org/abs/1203.3544}{{\ttfamily
  1203.3544}}].

\bibitem{Jensen:2012jh}
K.~Jensen, M.~Kaminski, P.~Kovtun, R.~Meyer, A.~Ritz and A.~Yarom,
  \emph{{Towards hydrodynamics without an entropy current}},
  \href{https://doi.org/10.1103/PhysRevLett.109.101601}{\emph{Phys. Rev. Lett.}
  {\bfseries 109} (2012) 101601}
  [\href{https://arxiv.org/abs/1203.3556}{{\ttfamily 1203.3556}}].

\bibitem{Baier:2007ix}
R.~Baier, P.~Romatschke, D.~T. Son, A.~O. Starinets and M.~A. Stephanov,
  \emph{{Relativistic viscous hydrodynamics, conformal invariance, and
  holography}},
  \href{https://doi.org/10.1088/1126-6708/2008/04/100}{\emph{JHEP} {\bfseries
  04} (2008) 100} [\href{https://arxiv.org/abs/0712.2451}{{\ttfamily
  0712.2451}}].

\bibitem{Romatschke:2009ng}
P.~Romatschke and D.~T. Son, \emph{{Spectral sum rules for the quark-gluon
  plasma}}, \href{https://doi.org/10.1103/PhysRevD.80.065021}{\emph{Phys. Rev.}
  {\bfseries D80} (2009) 065021}
  [\href{https://arxiv.org/abs/0903.3946}{{\ttfamily 0903.3946}}].

\bibitem{Moore:2012tc}
G.~D. Moore and K.~A. Sohrabi, \emph{{Thermodynamical second-order hydrodynamic
  coefficients}}, \href{https://doi.org/10.1007/JHEP11(2012)148}{\emph{JHEP}
  {\bfseries 11} (2012) 148} [\href{https://arxiv.org/abs/1210.3340}{{\ttfamily
  1210.3340}}].

\bibitem{Kovtun:2018dvd}
P.~Kovtun and A.~Shukla, \emph{{Kubo formulas for thermodynamic transport
  coefficients}}, \href{https://doi.org/10.1007/JHEP10(2018)007}{\emph{JHEP}
  {\bfseries 10} (2018) 007}
  [\href{https://arxiv.org/abs/1806.05774}{{\ttfamily 1806.05774}}].

\bibitem{Maxwell}
J.~C. Maxwell, \emph{The Theory of Heat}. Longmans, Green, and Co., 1871.

\bibitem{Nakazawa:1984zq}
N.~Nakazawa and T.~Fukuyama, \emph{{On the Energy Momentum Tensor at Finite
  Temperature in Curved Space-time}},
  \href{https://doi.org/10.1016/0550-3213(85)90465-1}{\emph{Nucl. Phys. B}
  {\bfseries 252} (1985) 621}.

\bibitem{Dowker:1988jw}
J.~S. Dowker and J.~P. Schofield, \emph{{High Temperature Expansion of the Free
  Energy of a Massive Scalar Field in a Curved Space}},
  \href{https://doi.org/10.1103/PhysRevD.38.3327}{\emph{Phys. Rev. D}
  {\bfseries 38} (1988) 3327}.

\bibitem{Gribosky:1988yk}
P.~S. Gribosky, J.~F. Donoghue and B.~R. Holstein, \emph{{The Stability of Hot
  Curved Space}},
  \href{https://doi.org/10.1016/0003-4916(89)90263-7}{\emph{Annals Phys.}
  {\bfseries 190} (1989) 149}.

\bibitem{Bhattacharyya:2008jc}
S.~Bhattacharyya, V.~E. Hubeny, S.~Minwalla and M.~Rangamani, \emph{{Nonlinear
  Fluid Dynamics from Gravity}},
  \href{https://doi.org/10.1088/1126-6708/2008/02/045}{\emph{JHEP} {\bfseries
  02} (2008) 045} [\href{https://arxiv.org/abs/0712.2456}{{\ttfamily
  0712.2456}}].

\bibitem{Jensen:2011xb}
K.~Jensen, M.~Kaminski, P.~Kovtun, R.~Meyer, A.~Ritz and A.~Yarom,
  \emph{{Parity-Violating Hydrodynamics in 2+1 Dimensions}},
  \href{https://doi.org/10.1007/JHEP05(2012)102}{\emph{JHEP} {\bfseries 05}
  (2012) 102} [\href{https://arxiv.org/abs/1112.4498}{{\ttfamily 1112.4498}}].

\bibitem{Moore:2010bu}
G.~D. Moore and K.~A. Sohrabi, \emph{{Kubo Formulae for Second-Order
  Hydrodynamic Coefficients}},
  \href{https://doi.org/10.1103/PhysRevLett.106.122302}{\emph{Phys. Rev. Lett.}
  {\bfseries 106} (2011) 122302}
  [\href{https://arxiv.org/abs/1007.5333}{{\ttfamily 1007.5333}}].

\bibitem{Megias:2014mba}
E.~Megias and M.~Valle, \emph{{Second-order partition function of a
  non-interacting chiral fluid in 3+1 dimensions}},
  \href{https://doi.org/10.1007/JHEP11(2014)005}{\emph{JHEP} {\bfseries 11}
  (2014) 005} [\href{https://arxiv.org/abs/1408.0165}{{\ttfamily 1408.0165}}].

\bibitem{Buzzegoli:2017cqy}
M.~Buzzegoli, E.~Grossi and F.~Becattini, \emph{{General equilibrium
  second-order hydrodynamic coefficients for free quantum fields}},
  \href{https://doi.org/10.1007/JHEP10(2017)091}{\emph{JHEP} {\bfseries 10}
  (2017) 091} [\href{https://arxiv.org/abs/1704.02808}{{\ttfamily
  1704.02808}}].

\bibitem{Philipsen:2013nea}
O.~Philipsen and C.~Sch{\"a}fer, \emph{{The second order hydrodynamic transport
  coefficient $\kappa$ for the gluon plasma from the lattice}},
  \href{https://doi.org/10.1007/JHEP02(2014)003}{\emph{JHEP} {\bfseries 02}
  (2014) 003} [\href{https://arxiv.org/abs/1311.6618}{{\ttfamily 1311.6618}}].

\bibitem{Romatschke:2019gck}
P.~Romatschke, \emph{{Analytic Transport from Weak to Strong Coupling in the
  O(N) model}}, \href{https://doi.org/10.1103/PhysRevD.100.054029}{\emph{Phys.
  Rev. D} {\bfseries 100} (2019) 054029}
  [\href{https://arxiv.org/abs/1905.09290}{{\ttfamily 1905.09290}}].

\bibitem{Lawrence:2022vwa}
S.~Lawrence and P.~Romatschke, \emph{{Gravitational-wave-to-matter coupling of
  superfluid Fermi gases near unitarity}},
  \href{https://doi.org/10.1103/PhysRevA.107.033327}{\emph{Phys. Rev. A}
  {\bfseries 107} (2023) 033327}
  [\href{https://arxiv.org/abs/2206.04765}{{\ttfamily 2206.04765}}].

\bibitem{Buchel:2008bz}
A.~Buchel and M.~Paulos, \emph{{Relaxation time of a CFT plasma at finite
  coupling}},
  \href{https://doi.org/10.1016/j.nuclphysb.2008.07.002}{\emph{Nucl. Phys. B}
  {\bfseries 805} (2008) 59} [\href{https://arxiv.org/abs/0806.0788}{{\ttfamily
  0806.0788}}].

\bibitem{Arnold:2011ja}
P.~Arnold, D.~Vaman, C.~Wu and W.~Xiao, \emph{{Second order hydrodynamic
  coefficients from 3-point stress tensor correlators via AdS/CFT}},
  \href{https://doi.org/10.1007/JHEP10(2011)033}{\emph{JHEP} {\bfseries 10}
  (2011) 033} [\href{https://arxiv.org/abs/1105.4645}{{\ttfamily 1105.4645}}].

\bibitem{PhysRev.58.919}
C.~Eckart, \emph{The thermodynamics of irreversible processes. {III.}
  {R}elativistic theory of the simple fluid},
  \href{https://doi.org/10.1103/PhysRev.58.919}{\emph{Phys. Rev.} {\bfseries
  58} (1940) 919}.

\bibitem{weinberg:1972}
{S.~Weinberg}, \emph{{Gravitation and Cosmology}}. {John Wiley \& Sons},
  {1972}.

\end{thebibliography}\endgroup

\end{document}